\documentclass[10pt,conference]{IEEEtran}
\IEEEoverridecommandlockouts
\usepackage{cite}
\usepackage{amsmath,amssymb,amsfonts}
\usepackage{algorithmic}
\usepackage{graphicx}
\usepackage{textcomp}
\usepackage{xcolor}

\usepackage{url}
\usepackage{comment}
\usepackage{booktabs}
\usepackage{stfloats}
\usepackage[T1]{fontenc}
\usepackage{tabularx}
\usepackage{pifont}
\usepackage{tikz}
\usepackage{hyperref}
\usepackage{orcidlink}
\usetikzlibrary{arrows.meta,positioning,calc,patterns}
\usepackage[table]{xcolor}
\DeclareMathSizes{5.7}{6}{5}{5}
\DeclareMathSizes{5.8}{6}{5}{5}
\DeclareMathSizes{6.0}{6}{5}{5}
\DeclareMathSizes{7.1}{7}{5}{5}

\newif\ifanonymous
\anonymousfalse %
\newcommand{\anon}[2]{\ifanonymous #1\else #2\fi}

\definecolor{draftmark}{RGB}{230,70,0}

\newcommand{\jqwik}{jqwik}
\newcommand{\se}{SE}
\newcommand{\evosuite}{EvoSuite}
\newcommand{\eqbench}{EqBench}
\newcommand{\togbench}{TOGBench}
\newcommand{\teralizer}{Teralizer}
\newcommand{\toolname}{EvoSuite-PBT} %
\newcommand{\rqanswer}[1]{\par\smallskip\noindent\fbox{\begin{minipage}{\dimexpr\columnwidth-2\fboxsep-2\fboxrule\relax}#1\end{minipage}}\par\smallskip}
\newcommand{\toomanyfilter}{\texttt{TooManyFilterMissesException} exception}

\definecolor{faulthl}{RGB}{255,224,224}
\definecolor{delhl}{RGB}{242,242,242}
\definecolor{okhl}{RGB}{222,244,226}
\definecolor{acchl}{RGB}{220,235,250}
\definecolor{orchl}{RGB}{254,240,214}
\tikzset{
  paperfig/.style={x=1pt, y=1pt, font=\sffamily,
    execute at begin picture={\setlength{\fboxsep}{1.2pt}},
    panel/.style={draw=black!30, line width=0.5pt, rounded corners=2.5pt, fill=white},
    subbox/.style={draw=black!22, line width=0.4pt, rounded corners=1.5pt, fill=black!2},
    num/.style={circle, fill=black, text=white, inner sep=0pt, minimum size=8.2pt,
      font=\sffamily\bfseries\fontsize{5.8}{6}\selectfont},
    ttl/.style={anchor=base west, yshift=-2.5pt, inner sep=0pt, font=\sffamily\bfseries\fontsize{7.1}{8.1}\selectfont},
    code/.style={anchor=west, inner sep=0pt, font=\fontfamily{lmtt}\fontsize{5.8}{6.9}\selectfont},
    txt/.style={anchor=west, inner sep=0pt, font=\sffamily\fontsize{6.0}{7.1}\selectfont},
    tag/.style={anchor=west, inner sep=0pt, font=\sffamily\fontsize{5.7}{6.7}\selectfont},
    hdr/.style={txt, font=\sffamily\bfseries\fontsize{5.7}{6.7}\selectfont, text=black!65},
    badge/.style={tag, line width=0.3pt, rounded corners=1.2pt, inner xsep=2.5pt, inner ysep=1.2pt},
    prebadge/.style={badge, fill=okhl, draw=green!45!black!40, text=green!35!black},
    postbadge/.style={badge, fill=orchl, draw=orange!60!black!40, text=orange!45!black},
    flow/.style={-{Triangle[length=3.0pt, width=4.6pt]}, line width=1.0pt, black!60},
    ptr/.style={-{Triangle[length=2.4pt, width=3.6pt]}, line width=0.55pt, black!45}
  }
}

\def\BibTeX{{\rm B\kern-.05em{\sc i\kern-.025em b}\kern-.08em
    T\kern-.1667em\lower.7ex\hbox{E}\kern-.125emX}}
\begin{document}

\title{Generating Property-Based Regression Tests from Source Code Semantics
}

\ifanonymous
\author{\and
\IEEEauthorblockN{Anonymous Author(s)}
\IEEEauthorblockA{\textit{Anonymous Department} \\
\textit{Anonymous Institution}\\
City, Country \\
email@anonymous.org}
\and
}
\else
\author{\IEEEauthorblockN{Clemens Bauer\,\orcidlink{0009-0000-9199-8563}}
\IEEEauthorblockA{\textit{Department of Informatics Systems} \\
\textit{University of Klagenfurt}\\
Klagenfurt, Austria \\
clemens.bauer@aau.at}
\and
\IEEEauthorblockN{Johann Glock\,\orcidlink{0000-0002-0152-8611}}
\IEEEauthorblockA{\textit{Software Competence Center Hagenberg GmbH} \\
Hagenberg, Austria \\
johann.glock@scch.at}
\and
\IEEEauthorblockN{Martin Pinzger\,\orcidlink{0000-0002-5536-3859}}
\IEEEauthorblockA{\textit{Department of Informatics Systems} \\
\textit{University of Klagenfurt}\\
Klagenfurt, Austria \\
martin.pinzger@aau.at}
}
\fi

\maketitle

\begin{abstract}
Conventional unit tests (CUTs) check a method under test at fixed concrete inputs. A test can therefore miss a regression if its input produces the expected output despite a fault. Property-based testing (PBT) checks universally quantified specifications across many randomly generated inputs, but writing properties and input generators by hand is difficult. Existing approaches automate PBT creation by generalizing an existing or generated unit test suite. They either over-approximate concrete inputs across repetitive tests or apply single-path symbolic analysis along concrete test runs.

This paper presents an approach that derives property-based regression tests directly from source code semantics via multi-path symbolic execution (SE). For each explored path through a target method, the approach records a path semantics comprising the path condition, the symbolic return value, and the state change map. Each path semantics induces a Hoare triple that compiles into an executable property. Because SE leaves parts of a class unspecified, the approach complements the generated properties with search-based CUTs for the whole class.

We evaluate our Java implementation against EvoSuite and the test generalization tool Teralizer on EqBench and on 11 projects from TOGBench. On EqBench, our approach keeps validated properties for 82.0\% of all methods. Its mean mutation score of 63.8\% exceeds the 55.2\% of EvoSuite and the 60.3\% of Teralizer. On TOGBench, target selection, SE limitations, and property validation restrict validated properties to 2.8\% of all methods. The resulting mean mutation score of 48.6\% falls below the 50.9\% of EvoSuite and the 51.1\% of Teralizer. Failures during test generation leave a few classes without effective tests and lead to this gap.
\end{abstract}


\section{Introduction}
\label{sec:introduction}
\begin{figure}[t]
\centering
\begin{tikzpicture}[paperfig]
\draw[rounded corners=3.5pt, fill=black!3, draw=black!20, line width=0.4pt] (0,0) rectangle (252,-348);

\draw[panel] (4,-4) rectangle (248,-134);
\node[num] at (12,-12) {1};
\node[ttl] at (19,-12) {Target method, boundary mutant, and point-witness CUTs};

\draw[subbox] (8,-21) rectangle (121,-130);
\node[hdr] at (11.5,-27) {Method under test};
\node[code] at (11.5,-36.5) {class BankAccount \{};
\node[code] at (11.5,-44.5) {\ \ double balance;};
\node[code] at (11.5,-52.5) {\ \ \dots};
\node[code] at (11.5,-60.5) {\ \ double withdraw(double amount) \{};
\fill[delhl] (9.5,-72.3) rectangle (119.5,-64.7);
\node[code, text=black!55] at (11.5,-68.5) {-\ \ \ if (amount > balance)};
\fill[faulthl] (9.5,-80.3) rectangle (119.5,-72.7);
\node[code, text=red!65!black] at (11.5,-76.5) {+\ \ \ if (amount >= balance)};
\node[code] at (11.5,-84.5) {\ \ \ \ \ \ return balance;};
\node[code] at (11.5,-92.5) {\ \ \ \ balance -= amount;};
\node[code] at (11.5,-100.5) {\ \ \ \ return balance;};
\node[code] at (11.5,-108.5) {\ \ \}};
\node[code] at (11.5,-116.5) {\}};

\draw[subbox] (125,-21) rectangle (244,-130);
\node[hdr] at (128.5,-27) {Generated CUTs (100\% branch coverage)};
\node[code] at (128.5,-36.5) {@Test void t0() \{};
\node[code] at (128.5,-44.5) {\ \ var a = new BankAccount(100);};
\node[code] at (128.5,-52.5) {\ \ assertEquals(100, a.withdraw(150));};
\node[code] at (128.5,-60.5) {\}};
\node[code] at (128.5,-68.5) {@Test void t1() \{};
\node[code] at (128.5,-76.5) {\ \ var a = new BankAccount(100);};
\node[code] at (128.5,-84.5) {\ \ assertEquals(70, a.withdraw(30));};
\node[code] at (128.5,-92.5) {\}};
\draw[black!18, line width=0.3pt] (128.5,-99.5) -- (240.5,-99.5);
\node[txt, text=red!65!black] at (128.5,-106.5) {Mutant \texttt{>=} survives: \texttt{t0} and \texttt{t1}};
\node[txt, text=black!65] at (128.5,-115.0) {execute the mutated branch but still pass};
\node[txt, text=black!65] at (128.5,-123.5) {by coincidental correctness.};

\draw[flow] (126,-134.8) -- (126,-141.7);

\draw[panel] (4,-142.5) rectangle (248,-225);
\node[num] at (12,-150.5) {2};
\node[ttl] at (19,-150.5) {Extracted path semantics $\langle \pi, \rho, \sigma \rangle$ and induced Hoare triple};

\node[anchor=north west, inner sep=0pt, font=\sffamily\fontsize{6.0}{7.2}\selectfont] at (8,-157.5) {%
  \renewcommand{\arraystretch}{1.18}%
  \arrayrulecolor{black!25}%
  \setlength{\arrayrulewidth}{0.35pt}%
  \setlength{\tabcolsep}{2pt}%
  \medmuskip=2mu \thickmuskip=3mu%
  \begin{tabular*}{236pt}{@{\hspace{2pt}}l@{\extracolsep{\fill}}l l l@{\hspace{2pt}}}
    \hline
    \multicolumn{1}{@{\hspace{2pt}}l}{\textcolor{black!65}{\textbf{\fontsize{5.7}{6.7}\selectfont Path}}} &
    \multicolumn{1}{l}{\textcolor{black!65}{\textbf{\fontsize{5.7}{6.7}\selectfont $\pi$}}} &
    \multicolumn{1}{l}{\textcolor{black!65}{\textbf{\fontsize{5.7}{6.7}\selectfont $\rho$}}} &
    \multicolumn{1}{l@{\hspace{2pt}}}{\textcolor{black!65}{\textbf{\fontsize{5.7}{6.7}\selectfont $\sigma$}}} \\
    \hline
    $p_1$ & $\mathit{amount} > \mathit{balance}$ & $\mathit{balance}$ & $\emptyset$ \\
    $p_{2.a}$ & $\mathit{amount} < \mathit{balance}$ & $\mathit{balance} - \mathit{amount}$ & $\{\mathit{balance} \mapsto \mathit{balance} - \mathit{amount}\}$ \\
    $p_{2.b}$ & $\mathit{amount} = \mathit{balance}$ & $\mathit{balance} - \mathit{amount}$ & $\{\mathit{balance} \mapsto \mathit{balance} - \mathit{amount}\}$ \\
    \hline
  \end{tabular*}%
};
\node[tag, text=black!50] at (10,-198) {SPF splits the \texttt{double} comparison into three states, two of which follow path $p_2$.};
\node[txt] at (10,-208) {Triple for $p_{2.b}$:\ \ $\{\mathit{amount} = a \wedge \mathit{balance} = b \wedge a = b\}$\ \ $r := \mathtt{withdraw}(\mathit{amount})$};
\node[txt] at (54.5,-217) {$\{r = b - a \wedge \mathit{balance} = b - a\}$};

\draw[flow] (126,-225.8) -- (126,-232.7);

\draw[panel] (4,-233.5) rectangle (248,-344);
\node[num] at (12,-241.5) {3};
\node[ttl] at (19,-241.5) {Compiled property for $p_{2.b}$};

\draw[subbox] (8,-250.5) rectangle (244,-340);

\node[code] at (11.5,-257.5) {@Property};
\node[code] at (11.5,-266) {void test2(@ForAll("gen2") InputVector v) \{};
\node[code] at (11.5,-274.5) {\ \ var acc = v.get(0); var a = v.get(1);};
\node[code] at (11.5,-283) {\ \ var expected = \colorbox{orchl}{acc.balance - a};};
\node[code] at (11.5,-291.5) {\ \ \colorbox{orchl}{assertEquals(expected, acc.withdraw(a));}};
\node[code] at (11.5,-300) {\ \ \colorbox{orchl}{assertEquals(expected, acc.balance);}};
\node[code] at (11.5,-308.5) {\}};
\node[code] at (11.5,-317) {@Provide Arbitrary<InputVector> gen2() \{};
\node[code] at (11.5,-325.5) {\ \ return combine(doubles().map(d -> new BankAccount(d)), doubles())};
\node[code] at (11.5,-334) {\ \ \ \ .filter(\colorbox{okhl}{v -> v.get(1) == v.get(0).balance}); \}};
\node[postbadge, anchor=east] at (240,-283) {postcondition $Q_{2.b}$ (fails on the mutant)};
\node[prebadge, anchor=east] at (240,-317) {precondition $\pi_{2.b}$};
\end{tikzpicture}
\caption{Running example on \texttt{withdraw}. The point-witness CUTs achieve branch coverage but miss the boundary mutant \texttt{amount >= balance}.}
\label{fig:motivating_example}
\end{figure}
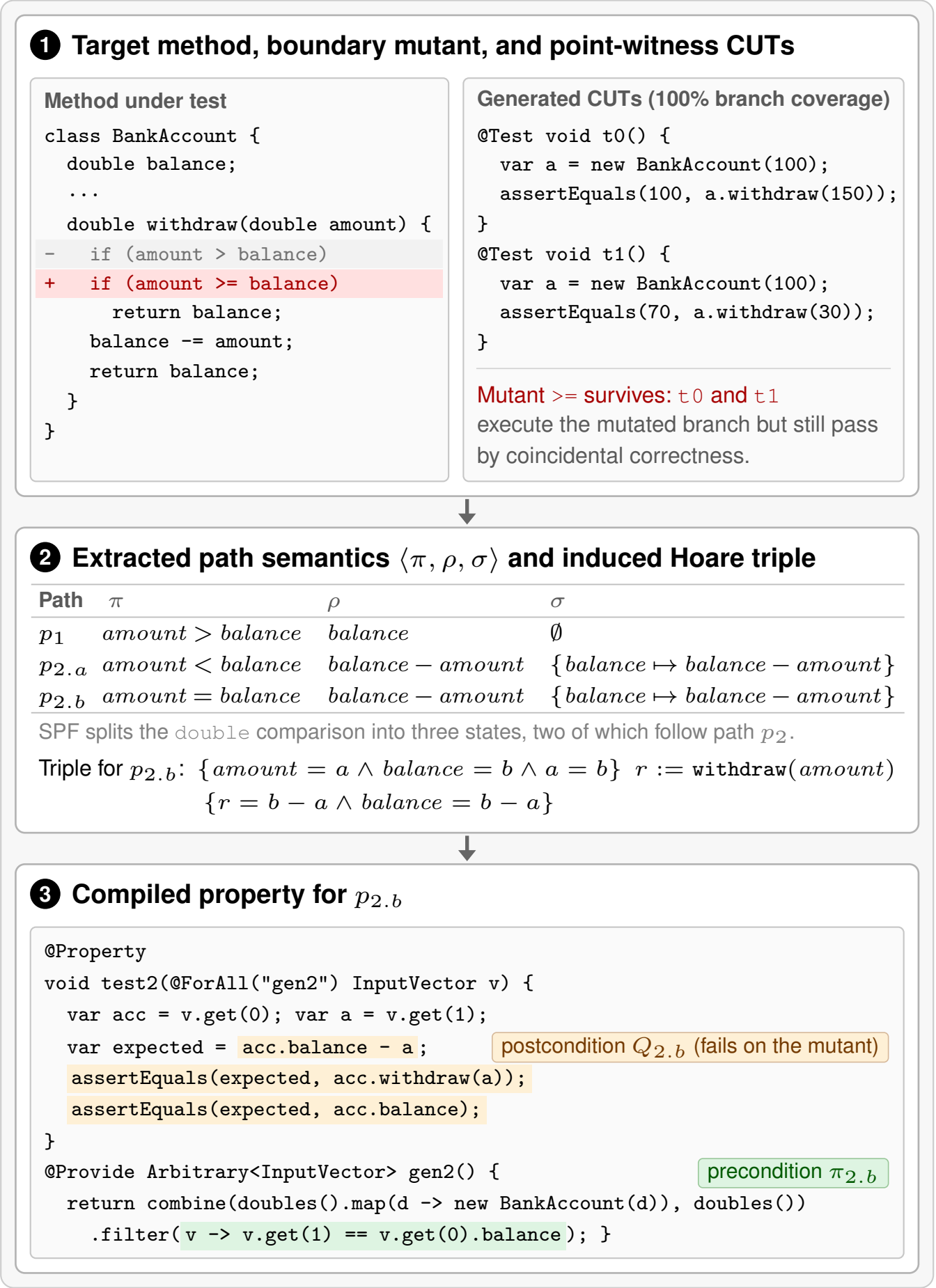

Automated unit test generators such as \evosuite{}~\cite{fraser2011evosuite} and Randoop~\cite{pacheco2007randoop} generate conventional unit tests (CUTs) that target structural coverage and assert the behavior observed at concrete inputs. High structural coverage, however, does not guarantee fault detection~\cite{inozemtseva2014effectiveness, hemmati2015weakCoverage}. When a change or mutant alters an operation along an executed path, a CUT still passes if the fault does not change the output for that test's input. This failure to expose an executed fault is known as coincidental correctness~\cite{voas1992pie, hierons2006coincidental}. Figure~\ref{fig:motivating_example}, Step~1 shows an example of a test suite containing CUTs that achieve full branch coverage, but fail to detect the behavioral change of the method under test. Data-driven test parameterization merges similar CUTs or recorded calls into parameterized unit tests (PUTs), but still checks only a fixed list of inputs~\cite{tsukamoto2018autoput, azamnouri2021compressingCUT, tiwari2024proze}. Property-based testing (PBT)~\cite{claessen2000quickcheck, hughes2007quickcheck} instead pairs an input generator with a parametric assertion and checks the assertion across newly sampled inputs on every run, killing more mutants per test than CUTs~\cite{ravi2025pbt}.

However, property-based tests remain rare in practice~\cite{ravi2025pbt, wauters2026evolution}. Developers struggle to state general postconditions without a reference implementation and to write generators that satisfy input preconditions~\cite{goldstein2024property, corgozinho2023pbt}. Manually converting existing unit tests into properties also requires rewriting test setup and oracle code~\cite{wauters2026evolution}. Recent approaches therefore automate PBT creation by generalizing existing or generated unit tests~\cite{peleg2018generatingPBTsByExample, glock2025teralizer}. Peleg et al.~\cite{peleg2018generatingPBTsByExample} group repetitive test execution traces into scenarios and instantiate predefined abstraction templates over their concrete values. Because these templates over-approximate the observed examples, the generated properties can fail on valid code. Glock et al.~\cite{glock2025teralizer} instead run single-path symbolic analysis along the concrete execution path of each CUT to extract a path-exact specification. However, starting from CUTs restricts generalization to execution paths that a CUT already traverses and to behaviors that its assertions explicitly check. Their tool \teralizer{} also requires mapping each assertion back to a single tested method call and supports only pure, non-void methods without field updates.

In this paper, we present \toolname{}, an approach that derives property-based regression tests from the source code of a target method rather than from its tests. \toolname{} explores the feasible paths through this method with multi-path symbolic execution (\se{})~\cite{SE_king, se_survey}.
Like symbolic method summaries in differential analysis~\cite{person2008differential}, our approach records the path semantics $\mathit{sem}(p) = \langle \pi, \rho, \sigma \rangle$ of each terminating path $p$ as the path condition $\pi$, the symbolic value $\rho$, and the state change map $\sigma$ over modified instance fields (Fig.~\ref{fig:motivating_example}, Step~2). When the \se{} engine models the path accurately, each path semantics induces a total-correctness Hoare triple~\cite{hoare1969axiomatic} that freezes pre-state values in logical variables $\bar{v}$ and specifies both the return value and post-state field updates. We compile each triple into an executable property.
Figure~\ref{fig:motivating_example}, Step~3 shows that the precondition $\pi_{2.b}$ becomes a generator \texttt{gen2} that filters for inputs with $\mathit{amount} = \mathit{balance}$. The postcondition becomes the assertions in \texttt{test2} on the return value $\rho(\bar{v})$ and the field $\sigma(\mathit{balance})(\bar{v})$.

Symbolic execution is incomplete in the presence of loops, unmodeled native calls, and complex objects~\cite{se_survey}, so the extracted properties form a partial specification of a class. Because filter-based generators can fail to sample enough valid inputs when a path condition is restrictive~\cite{claessen2000quickcheck, goldstein2024property}, our approach runs each candidate property for a small number of tries and discards properties that fail. It then complements the remaining properties with search-based CUTs generated for the remainder of the class.

We implemented our approach for Java on top of \evosuite{}~\cite{fraser2011evosuite} and its dynamic symbolic execution engine~\cite{galeotti2014evoDSE}, Symbolic PathFinder~\cite{puasuareanu2010symbolicPathfinder}, and \jqwik{}. We evaluated the implementation on the numeric benchmark \eqbench{}~\cite{badihi2021eqbench} and on 11 Java projects from \togbench{}~\cite{togbench2026} and compared it to the baselines \evosuite{} and \teralizer{}, the test generalization tool of Glock et al.~\cite{glock2025teralizer}.
On \eqbench{}, our approach kept validated properties for 82.0\% of all methods. Its mean mutation score of 63.8\% exceeded the 55.2\% of \evosuite{} and the 60.3\% of \teralizer{}. The mean test strength of 65.0\% likewise exceeded the 56.3\% of \evosuite{} and the 61.5\% of \teralizer{}. The properties account for this lead. On \togbench{}, target selection, \se{} limitations on object construction and constraint solving, and property validation restricted validated properties to 2.8\% of all methods. The mean mutation score of 48.6\% fell below the 50.9\% of \evosuite{} and the 51.1\% of \teralizer{}. Failures during test generation in a few classes led to this gap.
This paper makes the following contributions:
\begin{itemize}
    \item An approach that extracts symbolic path semantics $\langle \pi, \rho, \sigma \rangle$ over return values and instance-field updates via multi-path \se{} and compiles the induced Hoare triples into executable properties (Sections~\ref{sec:path_semantics}--\ref{sec:postcondition}).
    \item A hybrid test generation pipeline and open-source Java implementation, \toolname{}, that combines validated path-derived properties with whole-class search-based CUTs (Sections~\ref{sec:property_validation}--\ref{sec:implementation}).
    \item An empirical evaluation against \evosuite{} and \teralizer{} on \eqbench{} and \togbench{} measuring the fault-detection gains of direct extraction and the stage-by-stage applicability barriers on object-oriented code (Section~\ref{sec:evaluation}).
\end{itemize}
\section{Background}
This section reviews the two testing paradigms that our approach combines and the program analysis that connects them. It covers conventional unit testing, property-based testing with \jqwik{}, and symbolic execution.

\subsection{Conventional Unit Testing}
In conventional unit testing, a CUT invokes a target method in isolation with fixed arguments and compares the resulting return value or object state against an expected value~\cite{jorgensenSoftwareTestingCraftsmans2021}. Because a single input rarely exercises all branches of a method, test suites typically contain multiple CUTs per method. Repetitive CUTs that share the same assertion structure can be consolidated into a PUT, which runs the same test body across a variety of concrete input tuples supplied by a data source~\cite{tillmann2005parameterizedTests}. Although parameterization removes duplicate test code, the oracle remains tied to the finite set of concrete input-output examples listed in the test class.

\subsection{Property-Based Testing}
Property-based testing checks a method against a universally quantified specification across randomly generated inputs that satisfy a precondition~\cite{claessen2000quickcheck, goldstein2024property}. Our implementation targets \emph{\jqwik{}},\footnote{\url{https://jqwik.net}} a PBT engine for the Java Virtual Machine (JVM) that runs on the JUnit~5 platform. In \jqwik{}, a property is a method annotated with \texttt{@Property} whose parameters carry \texttt{@ForAll} annotations. Each parameter draws values from a type-specific generator, called an \texttt{Arbitrary}, which yields designated edge cases alongside random samples. Generators can be constrained declaratively through range annotations such as \texttt{@IntRange} or programmatically through provider methods annotated with \texttt{@Provide} that construct and filter \texttt{Arbitrary} instances. When a generator uses a filter predicate, \jqwik{} samples candidate inputs and discards those that violate the predicate. If too many consecutive candidates are rejected, \jqwik{} aborts the property with a \toomanyfilter{}.

\subsection{Symbolic Execution}
\label{sec:symbolic_execution}
Symbolic execution~\cite{SE_king, se_survey} analyzes a program on symbolic inputs that represent arbitrary values rather than concrete ones. Along an execution path, the \se{} engine maintains a symbolic state comprising a symbolic store, a path condition, and an instruction counter. The symbolic store maps program variables and object fields to symbolic expressions over the entry inputs, while the path condition conjoins the branch constraints accumulated along the path. At each conditional branch, the engine forks the symbolic state, conjoins the branch condition or its negation to the path condition, and queries a constraint solver to prune infeasible paths. When execution reaches a return instruction, the symbolic store and operand stack provide symbolic expressions for the return value and any modified instance fields.

In practice, \se{} is bounded by path explosion and constraint solving~\cite{se_survey}. Loops and nested branches can produce an exponential or infinite number of feasible paths, requiring bounds on search depth or input size, or symbolic state abstraction~\cite{jpf-se, se_survey}. Constraint solvers also face decidability and performance limits on floating-point arithmetic, strings, and nonlinear constraints, and standard \se{} cannot track execution across unmodeled native methods~\cite{jpf-se, se_survey}. Dynamic symbolic execution (DSE), or concolic execution, mitigates some of these limits by driving \se{} along concrete executions~\cite{se_survey, galeotti2013improving, galeotti2014evoDSE}. DSE runs the method on concrete inputs, collects the symbolic branch constraints along the executed path, and negates branch conditions to solve for new inputs that steer execution toward uncovered paths.

\section{Approach}
\label{sec:approach}
\begin{figure*}[t!]
\centering
\begin{tikzpicture}[paperfig,
  verdict/.style={anchor=center, inner sep=0pt, font=\fontsize{6.4}{7}\selectfont},
  note/.style={tag, text=black!50},
  sref/.style={tag, anchor=base east, yshift=-2.5pt, text=black!45},
  card/.style={draw=black!45, line width=0.4pt, rounded corners=0.8pt, fill=white},
  cardcode/.style={code, font=\fontfamily{lmtt}\fontsize{5.0}{6}\selectfont},
  edge/.style={line width=0.5pt, black!45},
  tdot/.style={circle, fill=black!55, inner sep=0pt, minimum size=2.6pt},
  trydot/.style={circle, fill=black!30, inner sep=0pt, minimum size=2.6pt},
  cutdot/.style={circle, fill=blue!55!black, draw=white, line width=0.3pt, inner sep=0pt, minimum size=3.2pt}
]
\draw[rounded corners=3.5pt, fill=black!3, draw=black!20, line width=0.4pt] (0,0) rectangle (516,-190);
\draw[flow] (168.8,-55) -- (175.2,-55);
\draw[flow] (340.8,-55) -- (347.2,-55);
\draw[flow] (86.5,-105.8) -- (86.5,-112.2);
\draw[flow] (429.5,-105.8) -- (429.5,-112.2);
\draw[flow] (347.2,-149) -- (340.8,-149);
\draw[flow] (168.8,-149) -- (175.2,-149);

\begin{scope}[shift={(5,-5)}]
\draw[panel] (0,0) rectangle (163,-100);
\node[num] at (8,-8) {1};
\node[ttl] at (15,-8) {Target selection};
\node[sref] at (158,-8) {\ref{sec:target_selection}};
\draw[card] (5,-16) rectangle (83,-94.6);
\fill[black!8, rounded corners=0.6pt] (5.3,-16.3) rectangle (82.7,-23.5);
\draw[black!35, line width=0.3pt] (5,-23.5) -- (83,-23.5);
\node[hdr] at (8,-19.9) {Class under test};
\fill[delhl] (5.6,-51.1) rectangle (82.4,-84.7);
\node[code] at (8,-30.1) {class C \{};
\node[code] at (8,-38.5) {\ \ static int f(int x)};
\node[code] at (8,-46.9) {\ \ void g(double d)};
\node[code, text=black!50] at (8,-55.3) {\ \ private int h(int x)};
\node[code, text=black!50] at (8,-63.7) {\ \ int k(Object o)};
\node[code, text=black!50] at (8,-72.1) {\ \ static long now()};
\node[code, text=black!50] at (8,-80.5) {\ \ String[] all()};
\node[code] at (8,-88.9) {\}};
\node[verdict, text=black!70] at (88,-38.5) {\ding{51}};
\node[verdict, text=black!70] at (88,-46.9) {\ding{51}};
\node[verdict, text=red!65!black] at (88,-55.3) {\ding{55}};
\node[verdict, text=red!65!black] at (88,-63.7) {\ding{55}};
\node[verdict, text=red!65!black] at (88,-72.1) {\ding{55}};
\node[verdict, text=red!65!black] at (88,-80.5) {\ding{55}};
\node[note] at (92,-55.3) {visibility};
\node[note] at (92,-63.7) {parameter type};
\node[note] at (92,-72.1) {parameter count};
\node[note] at (92,-80.5) {return type};
\end{scope}

\begin{scope}[shift={(176,-5)}]
\draw[panel] (0,0) rectangle (164,-100);
\node[num] at (8,-8) {2};
\node[ttl] at (15,-8) {Semantics extraction};
\node[sref] at (159,-8) {\ref{sec:extraction}};
\node[note] at (8,-19.5) {DSE for static methods, SPF for instance methods};
\node[tdot] (r) at (14,-63) {};
\node[tag, anchor=east] at (12,-63) {$m$};
\node[tdot] (a) at (27,-47) {};
\node[tdot] (b) at (27,-79) {};
\node[tdot] (l1) at (40,-39) {};
\node[tdot] (l2) at (40,-55) {};
\node[tdot] (l3) at (40,-71) {};
\draw[edge] (r) -- (a) (r) -- (b) (a) -- (l1) (a) -- (l2) (b) -- (l3);
\draw[edge, densely dashed] (b) -- (40,-87);
\node[txt] at (44,-39) {$p_1$: $\langle \pi_1, \rho_1, \sigma_1 \rangle$};
\node[txt] at (44,-55) {$p_2$: $\langle \pi_2, \rho_2, \sigma_2 \rangle$};
\node[txt] at (44,-71) {$p_3$: $\langle \pi_3, \rho_3, \sigma_3 \rangle$};
\node[note] at (44,-87) {not explored: native call};
\draw[black!15, line width=0.4pt] (98,-33) -- (98,-76);
\node[hdr] at (103,-39) {Read at each return};
\node[tag] at (103,-49) {$\pi$\ \ path condition};
\node[tag] at (103,-59) {$\rho$\ \ symbolic value};
\node[tag] at (103,-69) {$\sigma$\ \ state change map};
\end{scope}

\begin{scope}[shift={(348,-5)}]
\draw[panel] (0,0) rectangle (163,-100);
\node[num] at (8,-8) {3};
\node[ttl] at (15,-8) {Pre- and postcondition encoding};
\node[sref] at (158,-8) {\ref{sec:precondition}, \ref{sec:postcondition}};
\draw[card] (5,-16) rectangle (158,-94.6);
\fill[black!8, rounded corners=0.6pt] (5.3,-16.3) rectangle (157.7,-23.5);
\draw[black!35, line width=0.3pt] (5,-23.5) -- (158,-23.5);
\node[hdr] at (8,-19.9) {Candidate property for path $p_i$};
\node[code] at (8,-30.1) {@Property};
\node[code] at (8,-37.5) {void test$_i$(@ForAll("gen$_i$") InputVector v) \{};
\node[code] at (8,-44.9) {\ \ var expected = \kern-\fboxsep\colorbox{orchl}{$\rho_i$(v)}\kern-\fboxsep;};
\node[code] at (8,-52.3) {\ \ var expected$\ell$ = \kern-\fboxsep\colorbox{orchl}{$\sigma_i(\ell)$(v)}\kern-\fboxsep;};
\node[code] at (8,-59.7) {\ \ assertEquals(expected, m(v));};
\node[code] at (8,-67.1) {\ \ assertEquals(expected$\ell$, $\ell$);};
\node[code] at (8,-74.5) {\}};
\node[code] at (8,-81.9) {@Provide Arbitrary<InputVector> gen$_i$() \{};
\node[code] at (8,-89.3) {\ \ return combine(\dots).filter(\kern-\fboxsep\colorbox{okhl}{v -> $\pi_i$(v)}\kern-\fboxsep); \}};
\end{scope}

\begin{scope}[shift={(348,-113)}]
\draw[panel] (0,0) rectangle (163,-72);
\node[num] at (8,-8) {4};
\node[ttl] at (15,-8) {Property validation};
\node[sref] at (158,-8) {\ref{sec:property_validation}};
\node[note] at (8,-19.5) {10 tries per candidate property};
\node[txt] at (8,-30) {property for $p_1$};
\foreach \k in {0,...,9} \node[trydot] at (62+\k*4.6,-30) {};
\node[verdict, text=black!70] at (112,-30) {\ding{51}};
\node[txt, text=black!45] at (8,-41) {property for $p_2$};
\foreach \k in {0,...,2} \node[trydot] at (62+\k*4.6,-41) {};
\node[verdict, text=red!65!black] at (75.8,-41) {\ding{55}};
\node[note] at (80,-41) {too many filter misses};
\node[txt] at (8,-52) {property for $p_3$};
\foreach \k in {0,...,9} \node[trydot] at (62+\k*4.6,-52) {};
\node[verdict, text=black!70] at (112,-52) {\ding{51}};
\node[note] at (8,-63) {Failing properties are discarded.};
\end{scope}

\begin{scope}[shift={(176,-113)}]
\draw[panel] (0,0) rectangle (164,-72);
\node[num] at (8,-8) {5};
\node[ttl] at (15,-8) {Suite assembly};
\node[sref] at (159,-8) {\ref{sec:property_validation}};
\node[hdr] at (8,-19.5) {Input domain of $m$};
\begin{scope}[shift={(8,-24)}]
  \fill[okhl] (0,0) -- (26,0) -- (18,-42) -- (0,-42) -- cycle;
  \fill[white] (26,0) -- (50,0) -- (44,-42) -- (18,-42) -- cycle;
  \fill[okhl] (50,0) -- (72,0) -- (72,-24) -- (47.4,-24) -- cycle;
  \fill[pattern=north east lines, pattern color=black!25] (47.4,-24) -- (72,-24) -- (72,-42) -- (44,-42) -- cycle;
  \draw[black!35, line width=0.4pt] (26,0) -- (18,-42) (50,0) -- (44,-42) (47.4,-24) -- (72,-24);
  \draw[black!45, line width=0.5pt] (0,0) rectangle (72,-42);
  \node[tag, anchor=center, text=green!35!black] at (10,-21) {$p_1$};
  \node[tag, anchor=center, text=black!40] at (34,-21) {$p_2$};
  \node[tag, anchor=center, text=green!35!black] at (61,-12) {$p_3$};
  \foreach \x/\y in {5/-35, 14/-8, 30/-8, 37/-33, 57/-17, 60/-33, 67/-38} \node[cutdot] at (\x,\y) {};
\end{scope}
\node[hdr] at (88,-19.5) {Regression test suite};
\fill[okhl, draw=green!45!black!40, line width=0.3pt] (88,-25.8) rectangle (94,-31.2);
\node[tag] at (97,-28.5) {validated property};
\node[cutdot] at (91,-37) {};
\node[tag] at (97,-37) {CUT input};
\fill[white, draw=black!35, line width=0.3pt] (88,-42.8) rectangle (94,-48.2);
\node[tag] at (97,-45.5) {discarded property};
\fill[pattern=north east lines, pattern color=black!35] (88,-51.3) rectangle (94,-56.7);
\draw[black!35, line width=0.3pt] (88,-51.3) rectangle (94,-56.7);
\node[tag] at (97,-54) {unexplored path};
\end{scope}

\begin{scope}[shift={(5,-113)}]
\draw[panel] (0,0) rectangle (163,-72);
\node[ttl] at (8,-8) {Search-based CUT generation};
\node[sref] at (158,-8) {\ref{sec:property_validation}};
\foreach \d in {4,2} {
  \draw[card] (8+\d,-19+\d) rectangle (82+\d,-68+\d);
  \fill[acchl, rounded corners=0.6pt] (8.3+\d,-19.3+\d) rectangle (81.7+\d,-26.5+\d);
}
\draw[card] (8,-19) rectangle (82,-68);
\fill[acchl, rounded corners=0.6pt] (8.3,-19.3) rectangle (81.7,-26.5);
\draw[black!35, line width=0.3pt] (8,-26.5) -- (82,-26.5);
\node[hdr, text=blue!55!black] at (11,-22.9) {Generated CUTs};
\node[code] at (11,-33.1) {@Test};
\node[code] at (11,-40.5) {void t0() \{};
\node[code] at (11,-47.9) {\ \ int r = C.f(3);};
\node[code] at (11,-55.3) {\ \ assertEquals(3, r);};
\node[code] at (11,-62.7) {\}};
\node[hdr] at (94,-34) {\evosuite{}};
\node[tag] at (94,-43.5) {covers all methods of {\fontfamily{lmtt}\selectfont C},};
\node[tag] at (94,-52) {including excluded ones};
\end{scope}
\end{tikzpicture}
\caption{Overview of \toolname{}. Gray labels give the section that describes each step. Each explored path $p_i$ of a target method $m$ yields a path semantics $\langle \pi_i, \rho_i, \sigma_i \rangle$ and one candidate property, in which $\pi_i$, $\rho_i$, and $\sigma_i$ stand for their translations into Java expressions.}
\label{fig:overview}
\end{figure*}
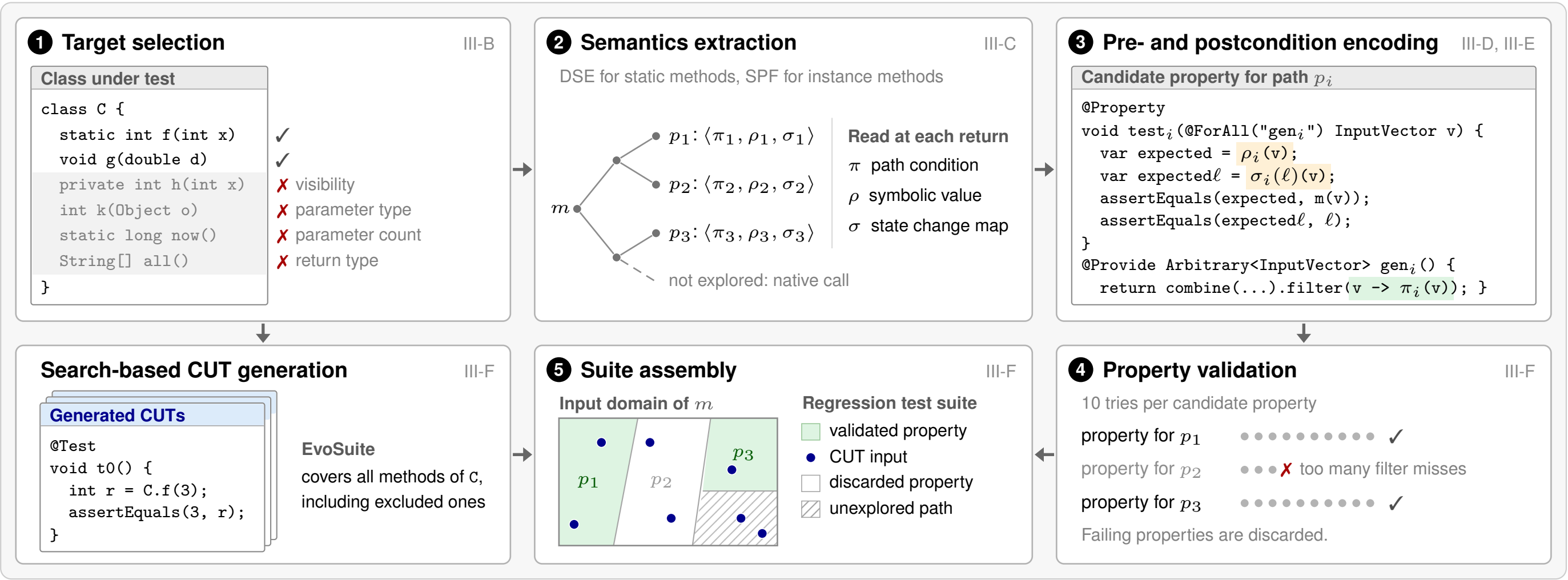

\toolname{} generates regression test suites that combine property-based tests derived from extracted program semantics with conventional unit tests. Rather than generalizing assertions and inputs from an existing or generated test suite~\cite{peleg2018generatingPBTsByExample, glock2025teralizer}, \toolname{} extracts the properties from the target methods via multi-path symbolic execution and complements them with fallback CUTs. For each explored path through a target method, the approach records a symbolic description of the method's behavior. Section~\ref{sec:path_semantics} formalizes this description as a Hoare triple.

Figure~\ref{fig:overview} shows how \toolname{} extracts these triples and compiles them into executable \jqwik{} properties. \emph{Target selection} (Section~\ref{sec:target_selection}) selects the methods for which the approach is actionable. \emph{Semantics extraction} (Section~\ref{sec:extraction}) computes one path semantics per explored path, and each path semantics yields one candidate property. \emph{Precondition encoding} (Section~\ref{sec:precondition}) compiles each path condition into a generator, and \emph{postcondition encoding} (Section~\ref{sec:postcondition}) compiles the symbolic value and the state update into the body of the property. The generator samples inputs that satisfy the path condition. The assertions compare the return value and the written fields with the symbolic value and the state update, both evaluated on the sampled inputs before the call.

Target selection and the limits of the \se{} engine leave parts of the class unspecified. \toolname{} therefore validates the candidate properties and complements them with CUTs that the search-based generator of \evosuite{} produces for the whole class. \emph{Property validation} (Section~\ref{sec:property_validation}) runs each candidate property for 10 tries and discards properties that fail. \emph{Suite assembly} (Section~\ref{sec:property_validation}) combines the surviving properties with the generated CUTs into a regression test suite of JUnit~5 and \jqwik{} sources. On the input domain of a target method, each validated property samples the input space of its path, whereas CUTs check only fixed concrete inputs.

\subsection{Path Semantics as Hoare Triples}
\label{sec:path_semantics}

Let $m$ be a method under test and $p$ a feasible path through $m$ that ends in a return. The \emph{symbolic inputs} $\bar{x} = (x_1, \dots, x_n)$ are the parameters of $m$ and, if $m$ is an instance method, any fields that it references. The \emph{path semantics} of $p$ is the triple $\mathit{sem}(p) = \langle \pi, \rho, \sigma \rangle$ with the following components:
\begin{itemize}
    \item The \emph{path condition} $\pi$ is the conjunction of the branch constraints over $\bar{x}$ that the \se{} engine collects along $p$.
    \item The \emph{symbolic value} $\rho$ is an expression over $\bar{x}$ that denotes the value that $m$ returns on $p$.
    \item The \emph{state change map} $\sigma$ is a finite partial map from $\bar{x}$ to expressions over $\bar{x}$. Its domain $\mathrm{dom}(\sigma) \subseteq \bar{x}$ is the subset of instance fields in $\bar{x}$ modified along $p$, and $\sigma(\ell)$ denotes the final value of field $\ell$.
\end{itemize}
For an expression $e$ over $\bar{x}$ and a tuple of values $\bar{v}$, we write $e(\bar{v})$ for the value of $e$ under the substitution $\bar{x} \mapsto \bar{v}$, and $\mathrm{Sat}(\pi) = \{\bar{v} \mid \pi(\bar{v})\}$ for the set of inputs that satisfy $\pi$.

A path semantics induces a Hoare triple for the analyzed version of $m$:
\begin{equation}
\label{eq:triple}
  \{\, \bar{x} = \bar{v} \,\wedge\, \pi(\bar{v}) \,\} \quad r := m(\bar{x}) \quad \{\, Q_p(\bar{v}, r) \,\}
\end{equation}
with the postcondition
\begin{equation}
\label{eq:post}
  Q_p(\bar{v}, r) \;\equiv\; r = \rho(\bar{v}) \;\wedge \bigwedge_{\ell \in \mathrm{dom}(\sigma)} \ell = \sigma(\ell)(\bar{v}),
\end{equation}
where the first conjunct is absent if $m$ does not return a value. The logical variables $\bar{v}$ freeze the pre-state. They are necessary because $p$ may write a field that it also reads, so that an occurrence of $\ell$ in the postcondition denotes the post-state value of the field, whereas $\sigma(\ell)(\bar{v})$ refers to the values at method entry.

For a pure static method, $\sigma = \emptyset$ on every path and $Q_p(\bar{v}, r)$ reduces to $r = \rho(\bar{v})$. Figure~\ref{fig:motivating_example} illustrates the general stateful case on the instance method \texttt{withdraw}. The symbolic inputs of \texttt{withdraw} are the parameter $\mathit{amount}$ and the field $\mathit{balance}$, which every path reads. Because \texttt{amount} and \texttt{balance} have type \texttt{double}, the JVM compiles the comparison $\mathit{amount} > \mathit{balance}$ to a floating-point comparison instruction, for which SPF creates three states ($>$, $<$, and $=$), which yields the three path semantics.

We generate one property for each path semantics. Given $\bar{v}=(a,b)$, triple~(\ref{eq:triple}) instantiates for path~$p_{2.a}$ to
\begin{multline*}
  \{\, \mathit{amount} = a \wedge \mathit{balance} = b \wedge a < b \,\} \\
  r := \mathtt{withdraw}(\mathit{amount})\\ \;\; \{\, r = b - a \,\wedge\, \mathit{balance} = b - a \,\}.
\end{multline*}
The triple of path~$p_{2.b}$ has the same postcondition, but its precondition contains $a = b$ instead of $a < b$ (Section~\ref{sec:precondition}). Keeping this case as a separate property tests the boundary $\mathit{amount} = \mathit{balance}$ explicitly, where off-by-one faults such as writing \texttt{>=} instead of \texttt{>} change the behavior.

Triple~(\ref{eq:triple}) holds in the sense of total correctness for the version of $m$ from which \toolname{} extracts it. Every input that satisfies $\pi$ follows $p$, $p$ ends in a return, and the symbolic state at that return denotes the returned value and the written fields. This argument rests on the assumption that the engine captures the semantics of $p$ without error, which does not hold if $m$ is non-deterministic or executes constructs that the engine does not model. If the assumption holds, each triple is valid independently of the others. The quality of the exploration instead determines how much of $m$ the triples specify. Because $m$ is deterministic, the path conditions of distinct paths are mutually exclusive, so the extracted triples partition the part of the input domain that they cover. Their conjunction forms a \emph{partial specification} of $m$, and the fallback CUTs cover the input regions left unspecified.

\subsection{Target Selection}
\label{sec:target_selection}

Target selection filters the methods of a class by signature eligibility, selecting the methods whose parameters and results fall within the types that \toolname{} can model and encode. \toolname{} selects a method if it satisfies four criteria:
\begin{itemize}
    \item \emph{Visibility.} The method is not private. Although reflection would make private methods reachable, developers conventionally test them through a calling public method~\cite{aniche2022effective}.
    \item \emph{Parameter types.} Every parameter is of a primitive type or of type \texttt{String}. These are limitations introduced by the use of \se{} and the underlying constraint solvers.
    \item \emph{Parameter count.} In the case of a static method, there must be at least one parameter present to start the \se{} analysis. 
    \item \emph{Return type.} The return type must be primitive or of type \texttt{String}. For instance methods, \texttt{void} is also allowed. 
\end{itemize}
These criteria limit property generation to a subset of the methods of a class. The search-based generator of \evosuite{} applies weaker selection criteria and also produces CUTs for the excluded methods. CUTs therefore cover both eligible and excluded methods, whereas properties complement the CUTs only for the methods that pass target selection.

\subsection{Semantics Extraction}
\label{sec:extraction}

\toolname{} extracts path semantics using the dynamic symbolic execution (DSE) engine of \evosuite{}~\cite{galeotti2014evoDSE} for static methods and Symbolic PathFinder (SPF)~\cite{puasuareanu2010symbolicPathfinder}, an extension of Java PathFinder (JPF), for instance methods. The selected engine records the branch constraints and, if applicable, the state changes along the executed path. \toolname{} hooks into this process with a listener at the return instructions of the target frame. At such a point, the listener reads $\pi$ from the path-condition collector, takes $\rho$ as the topmost expression on the symbolic operand stack, and obtains $\sigma$ from the entries of the symbolic store.

The limits of the \se{} (Section~\ref{sec:symbolic_execution}) make the exploration incomplete, which is why \toolname{} under-approximates the behavior of $m$.
First, loops produce a number of paths that grows with each iteration and may be unbounded, which is not feasible to explore in a short time frame. Second, the \se{} engines do not model every language construct faithfully. Calls to native methods are one example.

\subsection{Precondition Encoding}
\label{sec:precondition}

Precondition encoding compiles $\pi$ into a generator, so that \jqwik{} samples only inputs that follow the path. \toolname{} supports two encodings and selects between them based on the shape of $\pi$. A constraint is \emph{simple} if it has the form $x \bowtie c$ or $c \bowtie x$ with an input parameter $x$, a constant $c$, and \mbox{$\bowtie\, \in \{=, <, \le, >, \ge\}$}. Every other constraint is \emph{complex}, in particular a relational constraint such as $x_1 < x_2$, a disequality such as $x \neq 0$, and a nonlinear constraint over a single input.

This distinction is used to select one of two encodings. A path condition consisting of only simple constraints can be represented using \jqwik{} annotations and does not require additional helper code. Other path conditions require custom generator code, which is supplied using a helper method. 

\paragraph{Annotation-based encoding}
The goal of annotation-based encoding is to reduce the size of the generated test code. It requires a path condition that only consists of simple constraints. Based on this assumption, the path condition can be modeled using annotations on the property's parameters provided by \jqwik{}. For a symbolic input parameter $x$, let $C_{x}$ be the constraints of $\pi$ that mention $x$. Each simple constraint $c$ denotes an interval $I(c)$ over the domain of $x$, and \toolname{} computes the input space of $x$ as
\begin{equation}
\label{eq:interval}
  I(x) \;=\; \bigcap_{c \in C_x} I(c),
\end{equation}
starting from the full range of the declared type. The resulting interval is then modeled using the range annotations provided by \jqwik{}, such as \texttt{@IntRange} and \texttt{@FloatRange}. The path condition $\pi = x \ge 0$ for example compiles to \texttt{@IntRange(min = 0)}. An input parameter with $C_x = \emptyset$ remains unconstrained and receives a plain \texttt{@ForAll} annotation.

\paragraph{Provider-based encoding}
For every other path condition, \toolname{} emits a provider method annotated with \texttt{@Provide} that supplies an \texttt{Arbitrary} of \texttt{InputVector} objects. An \texttt{InputVector} is a container object that makes the encoding independent of the number and the types of the parameters. \toolname{} pushes the simple constraints of $\pi$ into the per-input arbitrary definitions as above and turns each complex constraint into a filter predicate. To obtain the predicate, it translates the symbolic constraint into a Java expression, replacing symbolic variables with references to the elements of the \texttt{InputVector}.

The provider-based generator is sound because \jqwik{} uses a sampled input only if it passes all filters. However, \jqwik{} samples first and then validates the input. If that input does not pass the filter, \jqwik{} rejects and resamples, so a restrictive $\pi$ raises the rejection rate and can abort the property with a \toomanyfilter{}. \toolname{} mitigates the most severe case, an equality constraint $x = c$, by generating \texttt{Arbitraries.of($c$)} instead of a filter, which reduces the rejection rate for that input to zero.
Handling more classes of complex constraints without filters would gradually turn the generator into a constraint solver~\cite{claessen2000quickcheck, claessen2015constrained, lampropoulos2017luck}. Solving $\pi$ on every try would increase test execution time, and a property whose solving exceeds the timeout could not be evaluated.

To create an arbitrary for the receiver object of an instance method, \toolname{} selects the constructor with the fewest parameters. \toolname{} draws the constructor arguments from \jqwik{} arbitraries and maps them through this constructor. However, this does not guarantee that the generator finds objects that satisfy the path condition, in which case the resulting properties fail with a \toomanyfilter{}.
Reflection could inject arbitraries directly into fields, but it can construct object states that violate the invariants of the class. Synthesizing object states that respect such unknown invariants is left for future work (Section~\ref{sec:conclusions}).
Figure~\ref{fig:motivating_example} (Step~3) shows the provider method for path~$p_{2.b}$ of \texttt{withdraw}, whose path condition is $\mathit{amount} = \mathit{balance}$. The provider maps a \texttt{double} arbitrary through the \texttt{BankAccount} constructor and pairs the resulting instance with the \texttt{amount} arbitrary inside an \texttt{InputVector}. Its filter predicate encodes the relational path condition and reads the \texttt{balance} field of the instance directly.
\toolname{} accesses non-private fields directly and falls back to getter methods when a field is \texttt{private}. If a \texttt{private} field has no getter, \toolname{} cannot map a constraint over this field, which causes the resulting property to fail.

\subsection{Postcondition Encoding}
\label{sec:postcondition}
Postcondition encoding compiles $Q_p$ into the assertion of the property. The property calls the target method on the generated inputs, which yields the actual return value, and evaluates $\rho$ on the same inputs, which yields the expected return value. For an instance method, the body additionally compares every field $\ell \in \mathrm{dom}(\sigma)$ after the call with $\sigma(\ell)$, checking the conjunction~(\ref{eq:post}) in full. Because the body evaluates $\rho$ and $\sigma(\ell)$ on the pre-call state rather than on the post-state, it realizes the logical freeze variables $\bar{v}$ of~(\ref{eq:triple}). Figure~\ref{fig:motivating_example} (Step~3) shows the generated property for path~$p_{2.b}$ of \texttt{withdraw}. The property evaluates $\rho_{2.b}$ before the call and stores the result in \texttt{expected}. Because $\sigma_{2.b}(\mathit{balance})$ equals $\rho_{2.b}$, the figure reuses \texttt{expected} as the expected post-state balance. The first assertion compares \texttt{expected} with the return value of \texttt{withdraw}, and the second compares it with the post-state \texttt{balance} field. The generated assertions compare \texttt{double} values with a \texttt{0.01} delta, which the figure omits.

The central ingredient in generating the postcondition statements is the translation from symbolic expressions to Java expressions, which \toolname{} implements as a visitor over the expression trees of the \se{} engine. The visitor resolves variable references according to the chosen precondition encoding by referencing either the property's parameters or the \texttt{InputVector} object. Arithmetic, bitwise, and comparison operators map to their Java counterparts, and the transcendental and rounding operators, which the engine represents in SMT-LIB style, map to the corresponding methods of \texttt{java.lang.Math}. For instance, \toolname{} translates the symbolic expression
  \texttt{\footnotesize (sqrt(var0) + 1.0)}
to
\[
  \texttt{\footnotesize ((Math.sqrt(arg0)) + 1.0)}
\]
under the annotation-based encoding, and to
\[
  \texttt{\footnotesize ((Math.sqrt(((double) inputVec.get(0)))) + 1.0)}
\]
under the provider-based encoding.

Two characteristics of this translation determine the accuracy of the oracle. First, the transformation is partial. It supports \texttt{String} as the only reference type and translates string operations such as \texttt{equals} and \texttt{charAt} to Java. Arrays and other reference types have no counterpart in the supported fragment, and the visitor signals this instead of emitting an approximation, dropping the current property. Second, the transformation reproduces the JDK operations that the method under test executes, so that the expected and the actual value result from the same numeric operations. The JDK specifies the transcendental functions of \texttt{java.lang.Math} only up to one or two units in the last place, so a property over such an expression inherits this tolerance and may behave differently across JVM implementations.

\subsection{Property Validation and Suite Assembly}
\label{sec:property_validation}
A compiled property can still fail to execute if its filter predicate rejects too many sampled inputs or if its generator cannot construct a valid input within the execution limits. Before emitting the final test suite, \toolname{} therefore validates each candidate property by executing it for 10 tries in \jqwik{}. Any property that fails during this check, for example by raising a \toomanyfilter{}, is discarded so that properties with broken generators do not enter the emitted test suite.

The properties that pass this validation check specify only the paths that \se{} explored within the targeted methods. To cover the remaining methods and paths of the class under test, \toolname{} runs the search-based generator of \evosuite{} for the whole class and combines the validated properties with the resulting CUTs into the final regression test suite.
\subsection{Implementation}
\label{sec:implementation}

We implemented \toolname{} on \evosuite{}~\cite{fraser2011evosuite}. We selected \evosuite{} over the other candidates of recent tool competitions~\cite{kifetew2025sbft} because of its performance, availability, and the fact that it emits JUnit~5 tests, which \jqwik{} requires. In addition to the DSE engine built into \evosuite{}, we added Symbolic PathFinder (SPF) for evaluating instance methods. \toolname{} takes compiled classes as input and emits JUnit~5 and \jqwik{} sources. These sources compile and run without \toolname{}. All runtime helpers needed for compiling and running the properties are provided within a separate dependency~\cite{evosuite_pbt_utils}.
We used this implementation of our approach for the evaluation described in the next section.
\section{Evaluation}
\label{sec:evaluation}
\begin{table}[t]
    \centering
    \caption{Benchmarks used in the evaluation. LOC denotes the lines of code as reported by \texttt{cloc}.}
    \label{tab:benchmarks}
    \footnotesize
    \begin{tabular}{@{}llr@{}}
        \toprule
        Name & Description & LOC \\
        \midrule
        \eqbench{} & Equivalent and non-equivalent & 23,939 \\
        & program pairs & \\
        \midrule
        JSON-java & JSON encoder and decoder & 4,195 \\
        async-http-client & Asynchronous HTTP client & 15,753 \\
        commons-dbutils & JDBC utility classes & 3,074 \\
        commons-geometry & Euclidean and spherical geometry & 15,246 \\
        commons-jcs & JCache caching implementation & 6,158 \\
        commons-numbers & Numerical utilities & 5,416 \\
        commons-validator & Data validation routines & 7,619 \\
        commons-vfs2-jackrabbit1 & WebDAV provider for  & 705 \\
        & Commons VFS & \\
        http-request & HTTP request convenience library & 1,391 \\
        scribejava-core & OAuth client library & 3,771 \\
        springside4-metrics & Application metrics library & 1,213 \\
        \midrule
        Total & & 88,480 \\
        \bottomrule
    \end{tabular}
\end{table}
We evaluated \toolname{} on two distinct datasets to answer the following research questions:
\vspace{3mm}

\textbf{RQ1 (Applicability).} To what extent can \toolname{} generate properties for the methods under test, and what limits it?

\vspace{1mm}

\textbf{RQ2 (Effectiveness).} How do the test suites that \toolname{} generates compare to those that \evosuite{} and \teralizer{} generate in terms of test effectiveness?

\vspace{3mm}
We evaluated \toolname{} first on \eqbench{}~\cite{badihi2021eqbench}, which is a synthetic dataset comprised of Java methods that perform arithmetic operations. Because \eqbench{} does not contain object-oriented code, we also evaluated \toolname{} on 11 projects from the \togbench{}~\cite{togbench2026} dataset. \togbench{} is a benchmark for automated test oracle generation built from developer-written unit tests in 25 real-world Java projects. We chose \togbench{} because it was published in 2026 and all of its projects target Java 8. We selected the 11 projects on which we could run all three tools. The generated regression test suites were compared against two baselines, \evosuite{}~\cite{fraser2011evosuite} and \teralizer{}~\cite{glock2025teralizer}. Table~\ref{tab:benchmarks} lists the selected projects and their size.

\evosuite{} serves as the first baseline, because \toolname{} builds on it (Section~\ref{sec:implementation}) and uses the same search-based generator for its CUTs. This comparison isolates the effect of property generation. \teralizer{} serves as the second baseline, because it also derives properties from program semantics. It generalizes existing CUTs into properties through single-path symbolic analysis. Each \teralizer{} run generalized and retained the CUTs of the corresponding \evosuite{} run. Comparing \teralizer{} with \evosuite{} thus shows what the generalized properties add to the fault detection of the same CUTs.

We ran \evosuite{} and \toolname{} with a search budget of 60~seconds per class and a fixed random seed. This budget covers the search phase of \toolname{}. The 10-try check runs afterward as a post-processing step (Section~\ref{sec:property_validation}) and does not count toward the search budget. We repeated each evaluation three times per dataset. We ran all evaluations on an Ubuntu 24.04.4 LTS x86\_64 virtual machine with an AMD EPYC 9655 4-core CPU at 2.5\,GHz and 16\,GB of RAM\@. For \teralizer{}, we enabled only its test generalization and configured it to use the \texttt{IMPROVED} algorithm. We then executed the generated properties of both tools for 200 \jqwik{} tries each, with a 4-second timeout. A replication package for this evaluation is available on Zenodo~\cite{anonymous_2026_23976317}.

\begin{table}[t]
    \centering
    \caption{Applicability of \toolname{} per dataset. Counts are means over the three runs.}
    \label{tab:applicability}
    \footnotesize
    \begin{tabular}{@{}lrr@{}}
        \toprule
        & \eqbench{} & \togbench{} \\
        \midrule
        Methods in total & 1,256 & 7,107 \\
        Methods selected for \se{} & 1,116 & 510 \\
        \se{} analysis successful & 1,116 & 385 \\
        Methods with at least one property & 1,062 & 274 \\
        Methods with at least one kept property & 1,030 & 198 \\
        \midrule
        Generated properties & 5,142 & 20,745 \\
        Properties discarded at the 10-try check & 1,046 & 20,260 \\
        Properties unexecutable during generation & 0 & 12 \\
        Properties remaining after generation & 4,096 & 474 \\
        Properties executed during the evaluation & 4,096 & 358 \\
        Properties failing during the evaluation & 860 & 25 \\
        \bottomrule
    \end{tabular}
\end{table}

\subsection{Applicability}
\label{sec:rq1}
\begin{figure}[t]
    \centering
    \includegraphics[width=\linewidth]{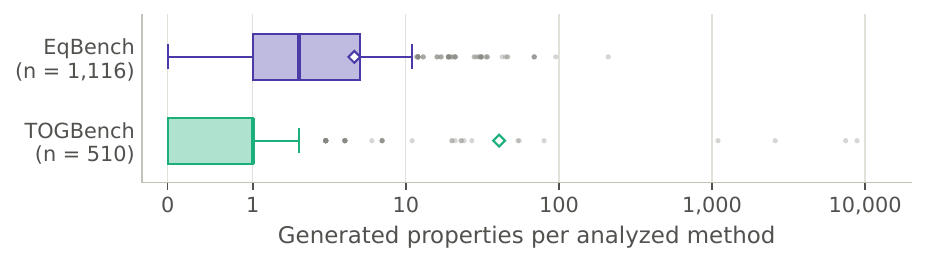}
    \caption{Generated properties per method analyzed by \se{}, including methods without properties, on a symmetric logarithmic scale with a linear threshold of 1. Counts are means over the three runs. The diamonds mark the mean.}
    \label{fig:properties_per_method}
\end{figure}

This section answers RQ1 by counting the methods for which \toolname{} generated at least one property and by classifying the reasons for which it generated none. Table~\ref{tab:applicability} reports these counts per dataset as means over the three runs. Figure~\ref{fig:properties_per_method} shows the distribution of generated properties per method analyzed by \se{}.

\paragraph{\eqbench{}} \toolname{} generated properties for most methods of \eqbench{}. Target selection (Section~\ref{sec:target_selection}) selected {1,116} of its {1,256} methods and excluded the other {140}, because they violate at least one of its four criteria. \se{} analyzed all targeted methods without error. \toolname{} generated at least one property for {1,062} analyzed methods, and \se{} found no paths for the other {54}.
After the 10-try check, at least one property remained for {1,030} analyzed methods, which corresponds to {82.0\%} of all methods.

The analyzed methods of \eqbench{} received a median of {2} properties.
Half of them received between {1} and {5} properties. On average, each analysis yielded {4.6} properties. The outliers, with up to {210} properties, are methods that contain loops, which lead to many paths through the method.

The 10-try check discarded one in five properties of \eqbench{}. \toolname{} generated {5,142} properties for \eqbench{}. During generation, the 10-try check discarded {1,046} of them. Most of these properties failed due to the \toomanyfilter{}. In the evaluation with 200 tries, {860} of the remaining {4,096} properties failed despite passing the 10-try check. Most of them exceeded the 4-second timeout.

\paragraph{\togbench{}} Target selection and \se{} restricted \toolname{} to a small share of the methods in the \togbench{} projects. Target selection selected {510} of their {7,107} methods and excluded the other {6,597}, because they violate at least one of its four criteria.
\se{} analyzed {385} targeted methods without error. On the other {125} methods, \se{} failed due to JPF crashes, constraint-solver limitations, or failures to generate the instrumentation harness. \toolname{} generated at least one property for {274} analyzed methods, and \se{} found no paths for the other {111}. After the 10-try check, at least one property remained for {198} analyzed methods, which corresponds to {2.8\%} of all methods.

Most analyzed methods of \togbench{} received at most one property. The median method received {1} property. Half of the methods received between {0} and {1} properties. The maximum for a single method is {9,386} properties. These outliers lift the mean to 40.7 properties per method analysis.

The 10-try check discarded almost all properties of \togbench{}. \toolname{} generated {20,745} properties for the \togbench{} projects. During generation, the 10-try check discarded {20,260} of them. {18,002} these properties failed due to invalid string generators. Another {12} properties could not be executed, because the \evosuite{} client thread crashed for their classes. These crashes left the classes without any tests. During the evaluation, {358} of the {474} remaining properties ran. Out-of-memory errors during the JUnit execution prevented the execution of the other {116} properties. Of the executed properties, {25} failed within the 200 tries despite passing the 10-try check, most of them by raising the \toomanyfilter{}. The remaining 333 properties passed the execution without errors.

\rqanswer{\textbf{Answer to RQ1.} \toolname{} generated at least one property for {95.2\%} of the analyzed methods of \eqbench{} and for {71.2\%} of the analyzed methods of the \togbench{} projects. After the 10-try check, properties remained for {92.3\%} and {51.4\%} of these methods, which corresponds to {82.0\%} and {2.8\%} of all methods. In the \togbench{} projects, target selection excluded {92.8\%} of the methods, because they violated at least one of its four criteria. \se{} failed on {24.5\%} of the targeted methods due to JPF crashes, constraint-solver limitations, or failures to generate the instrumentation harness. Of the generated properties, {97.7\%} failed the 10-try check, mostly due to invalid string generators.}

\subsection{Effectiveness}
\label{sec:rq2}
\begin{figure}
    \centering
    \includegraphics[width=\linewidth]{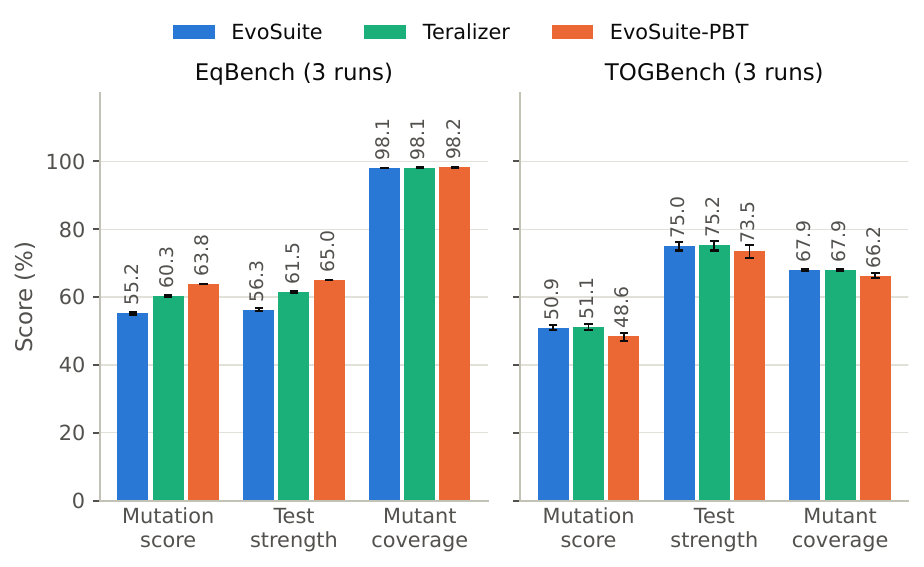}
    \caption{Mean mutation score, test strength, and mutant coverage of \evosuite{}, \teralizer{}, and \toolname{} for \eqbench{} and \togbench{}. The error bars span the minimum and maximum over the three runs.}
    \label{fig:overall_scores}
\end{figure}
This section answers RQ2 using mutation analysis~\cite{demillo1978hints, budd1978mutationsystem}. Mutation analysis injects small syntactic changes, called mutants, into a program and checks whether the test suite distinguishes each mutant from the original. It assumes that programmers write nearly correct programs and that tests detecting simple faults also detect complex ones, which is known as the coupling effect~\cite{demillo1978hints, budd1978mutationsystem}. We use PIT~\cite{coles2016pit} with its default mutators. PIT generates mutants deterministically, so all three tools face the same mutant set.
PIT first runs all tests on the unmutated bytecode, excludes tests that do not pass, and records which mutant each remaining test covers. It then runs each mutant only against the tests that cover it.
A remaining test kills a mutant if it changes its behavior, e.g. fails, times out, or exhausts or crashes the test JVM. The mutation score counts all generated mutants, including uncovered ones:
\begin{equation}
\label{eq:mutation_score}
  \mathit{MS} = \frac{\#\mathit{killed}}{\#\mathit{total}}.
\end{equation}
To measure fault detection on the exercised code alone, we also report PIT's test strength~\cite{coles2016pit}, which corresponds to the normalized effectiveness of Inozemtseva and Holmes~\cite{inozemtseva2014effectiveness}:
\begin{equation}
\label{eq:test_strength}
  \mathit{TS} = \frac{\#\mathit{killed}}{\#\mathit{covered}}.
\end{equation}

Figure~\ref{fig:overall_scores} shows the mutation score, test strength, and mutant coverage of the three tools on both datasets. The two baselines reached almost the same mutant coverage, because each \teralizer{} suite retains the CUTs of its \evosuite{} input.

\paragraph{\eqbench{}} \toolname{} generated more effective suites than both baselines on \eqbench{}. Its mean mutation score of 63.8\% and mean test strength of 65.0\% exceeded those of \evosuite{} by 8.6 and 8.7~percentage points. Over \teralizer{}, the gain amounted to 3.5~percentage points in both metrics.

These gains do not stem from broader coverage. On \eqbench{}, the mutant coverage of \toolname{} was 98.2\%, and that of both baselines was 98.1\%.
Because all three suites cover almost the same mutants, the higher test strength of \toolname{} means that its tests kill more of these mutants. The following counts show that its properties account for this difference.
We counted the mutants that every run of a tool killed but no run of \evosuite{} killed. \toolname{} killed 2,259 such mutants, 1.9 times as many as the 1,169 of \teralizer{}. In an average run, 2,188 of these mutants made a test of \toolname{} fail, and for 66.9\% of them the detecting test was a property. Properties of \toolname{} killed mutants in 328 of the 504 classes, whereas the generalized properties of \teralizer{} did so in 150 of these classes.

The lead of \toolname{} grows on the classes for which at least one of its properties passed the 10-try check.
On these 447 of the 504 classes, \toolname{} reached a mutation score of 66.5\%, against 57.4\% for \evosuite{} and 62.8\% for \teralizer{}.
Its lead thus grew to 9.2~percentage points over \evosuite{} and 3.7 over \teralizer{}. On the other 57 classes, all three tools reached a mutation score between 23.3\% and 23.4\%.

Per mutation operator, \toolname{} reached the highest mutation score on \eqbench{} for eight of the nine operators. Its largest gain occurred for the \texttt{Math} operator, which replaces arithmetic operators and produced 15,317 of the 23,961 mutants. For this operator, \toolname{} exceeded \evosuite{} by 10.8 and \teralizer{} by 3.7~percentage points. It scored lower than \evosuite{} only for the 160 \texttt{VoidMethodCall} mutants, with 1.0\% against 2.3\%.

The lead of \toolname{} on \eqbench{} lies entirely in static methods, which it explores with DSE (Section~\ref{sec:extraction}). Static methods contain 23,794 of the 23,961 mutants of \eqbench{}. On these mutants, \toolname{} reached a mutation score of 64.3\%, against 55.6\% for \evosuite{} and 60.7\% for \teralizer{}.

\paragraph{\togbench{}} \toolname{} was less effective than both baselines on the \togbench{} projects. Its mean mutation score of 48.6\% was 2.3~percentage points below that of \evosuite{} and 2.5~percentage points below that of \teralizer{}. Its mean test strength of 73.5\% was 1.5 and 1.8~percentage points lower, respectively. Its mutant coverage of 66.2\% fell 1.7~percentage points below the 67.9\% of both baselines. Compared to \teralizer{}, \toolname{} killed 494 fewer mutants per run, which corresponds to a relative decrease of 4.9\%.

Six classes account for this gap. They cost \toolname{} 494 killed mutants per run, more than its net deficit of 459 against \evosuite{}. Without them, \toolname{} exceeds \evosuite{} by 0.2~percentage points. For \texttt{JSON-java} and \texttt{commons-validator}, we had to raise the client timeout of \evosuite{} for \toolname{}, because it covers post-processing and some classes otherwise remained without tests. Even so, the logs show generation failures for five of the six classes. Test generation for \texttt{BigFraction} crashed in two of the three runs, and assertion generation for \texttt{ArithmeticUtils} timed out in all runs. The tests for \texttt{NtlmEngine} assert a random challenge, so they fail on the unmutated program and PIT excludes them. These failures affect CUTs, which \toolname{} generates in its own search, whereas each \teralizer{} run reuses the CUTs of the corresponding \evosuite{} run.

Properties contributed little on \togbench{}. Target selection, \se{} limitations, and the 10-try check restricted the remaining properties to 2.8\% of all methods (Section~\ref{sec:rq1}). As a result, properties of \toolname{} killed 93 mutants per run, spread over 21 of the 831 classes. The generalized properties of \teralizer{} reached fewer classes, killing 17 mutants per run in 9 classes. Still, \teralizer{} exceeded the mutation score of \evosuite{} by 0.2~percentage points, which corresponds to 35 additional killed mutants per run.

The deficit of \toolname{} on \togbench{} arises mainly in classes without a kept property. \toolname{} kept a property for 69 of the 831 classes, which contain 3,867 mutants. Because PIT mutates inner classes separately, the 496 mutants in the inner classes of these 69 classes count among the other 762 classes. On the 69 classes, the mutation score of \toolname{} was 63.5\%, which is 0.5~percentage points below that of \evosuite{} and 1.5 below that of \teralizer{}. On the other 762 classes, it fell 2.8~percentage points behind \evosuite{}.

On \togbench{}, \toolname{} scored lower than \evosuite{} for nine of the eleven operators. The two exceptions were \texttt{BooleanFalseReturnVals} with a gain of 1.5 and \texttt{ConditionalsBoundary} with a gain of 0.5~percentage points. Its largest loss occurred for \texttt{Math}, with 51.9\% against 59.4\% for \evosuite{} and 60.7\% for \teralizer{}. For all other operators, the mutation scores of \teralizer{} and \evosuite{} differed by at most 0.2~percentage points.

On \togbench{}, \toolname{} lost less on static methods, than on instance methods. \toolname{} scored 1.3~percentage points below \evosuite{} on static methods and 3.5~percentage points below on instance methods. A property was the first killing test for 2.9\% of the killed mutants in static methods and for 0.3\% of those in instance methods.

The comparison with \teralizer{} reflects a trade-off between extracting path semantics from the method under test and generalizing existing CUTs~\cite{glock2025teralizer}. Where \se{} can model a method, as on \eqbench{} and in \texttt{commons-numbers}, direct extraction reaches paths and boundary conditions that branch-covering CUTs do not traverse, and it needs no mapping from assertions to method calls. Test generalization instead reuses the valid receiver objects and concrete inputs of existing CUTs, which direct extraction must construct during \se{}~\cite{se_survey, jpf-se}.

\rqanswer{\textbf{Answer to RQ2.} On \eqbench{}, \toolname{} generated the most effective suites, with a mean mutation score of 63.8\% and a mean test strength of 65.0\%. These values exceed those of \evosuite{} by 8.6 and 8.7~percentage points and those of \teralizer{} by 3.5~percentage points each. Its properties account for this lead. On classes with a kept property, its lead over \evosuite{} grew to 9.2~percentage points. On \togbench{}, the mutation score of \toolname{} was 2.3 and 2.5~percentage points below that of \evosuite{} and \teralizer{}. Failures during test generation lead to this gap, because they left five of the six classes with the largest losses without effective tests. Properties remained for 2.8\% of all methods, so the deficit arises mainly in classes without a kept property.}

\section{Discussion}
\label{sec:discussion}
This section interprets the evaluation results and assesses how far they generalize. It derives implications for developers and researchers and then discusses threats to validity.

\subsection{Implications}
\label{sec:implications}
\paragraph{Implications for developers}
\toolname{} helps developers strengthen the regression test suites of methods that perform arithmetic operations. On \eqbench{} and the \texttt{gamma} and \texttt{combinatorics} modules of \texttt{commons-numbers} of \togbench{}, it raised the mutation score over \evosuite{} by 5.1 to 10.2~percentage points. Its properties derive from source code rather than from tests, so developers can add them to an existing suite or use them to test uncovered code.
Each property encodes the behavior of the current implementation, including its faults. Developers should therefore check each property against the intended behavior before they rely on it. This review spares them from writing postconditions and input generators by hand, which developers find difficult~\cite{goldstein2024property, corgozinho2023pbt}.

\paragraph{Implications for researchers}
On \eqbench{}, multi-path \se{} outperformed the single-path generalization of \teralizer{} by 3.5~percentage points, because it specifies paths that no CUT traverses. Research on property-based regression testing should therefore derive properties directly from source code wherever \se{} can model the method under test. On object-oriented code, the generation of valid objects limits this approach. On \togbench{}, target selection excluded 92.8\% of the methods, partly because it admits only primitive and \texttt{String} parameters. Of the generated properties, 97.7\% failed the 10-try check, mostly due to invalid string generators. Test generalization sidesteps object construction, because it starts from CUTs that already construct valid receiver objects. Direct extraction instead needs techniques that construct valid objects symbolically during \se{} and concretely in input generators. Research on such techniques will extend direct extraction to object-oriented code (Section~\ref{sec:conclusions}).

\subsection{Threats to Validity}
\label{sec:threats}

\paragraph{Internal validity}
Search-based test generation is time-bounded, so wall-clock scheduling introduces non-determinism across runs. To mitigate this threat, we repeated the evaluation three times per dataset for each tool using a fixed random seed and a 60-second limit per class, so the three runs differ only through wall-clock search timing.

\paragraph{External validity}
Our implementation of \toolname{} targets Java, which restricts our empirical findings to Java programs. The approach itself is applicable to other languages that provide a symbolic execution engine and a property-based testing framework. A further threat is that \eqbench{} is a synthetic dataset of arithmetic methods, so results on it may not generalize to real-world code. We mitigated this threat by also evaluating all three tools on 11 real-world Java projects from \togbench{}, which contain object-oriented code.

\paragraph{Construct validity}
A threat to construct validity is whether mutation score and test strength reflect fault-detection effectiveness on real faults. We mitigate this threat by evaluating with mutation analysis in PIT~\cite{coles2016pit}, whose mutants have been shown to reveal real faults in Java programs~\cite{kintis2018effectiveMutationTesting}.

\section{Related Work}
Prior work abstracts conventional unit tests into parameterized unit tests~\cite{tillmann2005parameterizedTests} by inferring symbolic constraints or by refactoring similar tests. Fraser and Zeller~\cite{fraser2011parameterizedTests} replace concrete inputs in generated CUTs with symbolic parameters and infer preconditions and postconditions through mutation analysis. AutoPUT~\cite{tsukamoto2018autoput} and the compression technique of Azamnouri and Paydar~\cite{azamnouri2021compressingCUT} instead merge similar CUTs into JUnit theories or parameterized tests over their literal values. Whereas these approaches depend on mutation analysis or on fixed literal values, our approach derives universally quantified properties directly from source code path semantics.

Other approaches construct parameterized tests from runtime execution traces rather than static test code alone. Tiwari et al.~\cite{tiwari2024proze} present PROZE, which instruments test executions to record runtime argument values for target methods, converts each single-assertion CUT into a JUnit~5 parameterized test backed by a method source over the observed argument tuples, and checks whether the existing assertion holds across those tuples. Kampmann and Zeller~\cite{kampmann2019carving} carve unit-level parameterized tests from system test runs by recording function arguments and global-variable writes, parameterizing arguments that appear as substrings of the system input, fuzzing those parameters at the unit level, and lifting inputs that cause crashes or new coverage back to the system level. Whereas PROZE replays a fixed set of recorded argument tuples against an existing assertion and system-test carving fuzzes carved parameters without a unit-level postcondition oracle, our approach derives both path-specific input generators and postcondition assertions from symbolic execution.

Specification mining approaches infer preconditions, postconditions, and object invariants from program executions rather than generating parameterized tests directly. Ernst et al.~\cite{ernst2007daikon} present Daikon, which instruments a program, observes variable values across a test suite, and reports candidate invariant templates that hold across the observed runs. Because template-based inference depends on the coverage of the test suite, sparse traces can produce overly restrictive invariants that reflect test-suite artifacts. Csallner et al.~\cite{csallner2008dysy} address this limitation in DySy by running symbolic execution alongside concrete test runs to derive path conditions and symbolic postconditions from the executed paths. However, both Daikon and DySy infer specifications only from executions of an existing test suite and do not compile inferred triples into executable property generators.

Recent work automates property-based test creation either from API documentation or by generalizing unit tests~\cite{vikram2023llmpbt, peleg2018generatingPBTsByExample, glock2025teralizer}. Vikram et al.~\cite{vikram2023llmpbt} prompt large language models with natural-language API documentation to synthesize generators and property assertions without existing unit tests, but model-generated tests can fail to compile or assert unsound properties when documentation is ambiguous. Peleg et al.~\cite{peleg2018generatingPBTsByExample} instead record concrete execution traces from repetitive CUTs, fold repeated call sequences, and instantiate predefined abstraction templates using an SMT solver. Because template instantiation over-approximates the observed examples, sparse example sets can yield properties that fail on valid inputs. Glock et al.~\cite{glock2025teralizer} generalize existing or generated CUTs into properties in \teralizer{} by running single-path symbolic analysis along the concrete path taken by each CUT. Because test generalization starts from CUTs, it only produces properties for execution paths that a CUT already traverses and for behaviors that its assertions check. \teralizer{} also requires mapping each test assertion to a single tested method call and supports only pure, non-void methods without field updates. In contrast, our approach extracts path semantics directly from the target method via multi-path symbolic execution and models instance-field updates in the induced Hoare triple.

\section{Conclusions}
\label{sec:conclusions}
This paper presented \toolname{}, an automated test generation approach that derives property-based regression tests directly from source code semantics via multi-path symbolic execution and combines them with search-based CUTs. For each explored path through a target method, \toolname{} records the path condition, symbolic return value, and instance-field state changes as a path semantics that induces a total-correctness Hoare triple. Each triple compiles into an executable \jqwik{} property whose precondition constrains input generation through range annotations or filter predicates and whose postcondition asserts both the return value and modified fields. Properties that pass a generation-time validation check are combined with whole-class CUTs generated by \evosuite{} for behavior that symbolic execution does not specify.

Evaluating \toolname{} against \evosuite{} and \teralizer{} on \eqbench{} and 11~\togbench{} projects showed that deriving properties directly from method path semantics can raise test effectiveness on methods that perform arithmetic operations. On \eqbench{}, the mean mutation score rose to 63.8\%, compared to 55.2\% for \evosuite{} and 60.3\% for \teralizer{}. On the numeric modules of \togbench{}, it rose by up to 10.2~percentage points over \evosuite{}. Across all 11~\togbench{} projects, however, target selection, \se{} limitations, and property validation restricted validated properties to 2.8\% of all methods.

For future work, we plan to address three limitations of \toolname{}. First, we plan to improve the synthesis of valid object states for reference parameters and fields~\cite{se_survey, jpf-se} to support methods beyond primitive and \texttt{String} signatures. Second, we will replace rejection sampling with constraint-aware generators~\cite{claessen2015constrained, lampropoulos2017luck} to reduce \toomanyfilter{} failures and evaluation timeouts. Third, we will apply test suite reduction and parameterization~\cite{tsukamoto2018autoput, azamnouri2021compressingCUT, glock2025teralizer} to merge the CUTs that validated properties make redundant.

\section*{Data Availability}
The replication package for this paper is available on Zenodo~\cite{anonymous_2026_23976317}. The runtime helpers that the generated properties require are available as a separate dependency~\cite{evosuite_pbt_utils}.

\section*{Acknowledgment}
\anon{This research was funded in whole or in part by an anonymized funding agency under an anonymized grant DOI.}{This research was funded in whole or in part by the Austrian Science Fund (FWF) 10.55776/P36698. For open access purposes, the author has applied a CC BY public copyright license to any author accepted manuscript version arising from this submission. The research reported in this paper has been partly funded by BMIMI, BMWET, and the State of Upper Austria in the frame of the SCCH competence center INTEGRATE (FFG grant no. 892418) part of the FFG COMET Competence Centers for Excellent Technologies Programme.} We used OpenAI GPT, Anthropic Claude, and Google Gemini models to assist with writing, coding, and literature search. The authors reviewed and refined the generated material and take full responsibility for the final manuscript.

\bibliographystyle{IEEEtran}
\bibliography{references}

\end{document}